\documentclass{aa}  

\usepackage{graphicx}
\usepackage{txfonts}
\usepackage{lipsum}
\usepackage{comment}
\usepackage{color}
\usepackage{subfigure}
\usepackage{capt-of}
\usepackage{hyperref}
\usepackage{multirow}
\usepackage[utf8]{inputenc}
\newcommand{\fink}{{\sc Fink}}

\newcommand{\referee}[1]{#1}
\newcommand{\review}[1]{#1}

\usepackage{array}
\newcommand\Tstrut{\rule{0pt}{2.6ex}}         % = `top' strut
\newcommand\Bstrut{\rule[-0.9ex]{0pt}{0pt}}   % = `bottom' strut

\begin{document}

   \title{Fink: early supernovae Ia classification using active learning}

   %\subtitle{I. Overviewing the $\kappa$-mechanism}

   \author{M. Leoni\inst{1}\thanks{\email{leoni@lal.in2p3.fr}},
          E. E. O. Ishida\inst{2}\thanks{\email{emille.ishida@clermont.in2p3.fr}},
          J. Peloton\inst{1} \and
         A. M\"oller\inst{2,3}
          }
   \institute{Université Paris-Saclay, CNRS/IN2P3, IJCLab, 91405 Orsay, France
    \and
             LPC, Université Clermont Auvergne, CNRS/IN2P3, F-63000 Clermont-Ferrand, France
   \and
   Centre for Astrophysics and Supercomputing, Swinburne University of Technology, Mail Number H29, PO Box 218, 31122 Hawthorn, VIC, Australia 
             }
   \date{Received September 15, 1996; accepted March 16, 1997}

% \abstract{}{}{}{}{} 
% 5 {} token are mandatory
 
  \abstract
   {The Vera C. Rubin Observatory Legacy  Survey of Space and Time (LSST) will produce a continuous stream of alerts made of varying sources in the sky. This data flow will be publicly advertised and distributed to scientists via broker systems such as \fink, whose task is to extract scientific information from the stream. Given the complexity and volume of the data to be generated, LSST is a prime target for machine learning (ML) techniques. One of the most challenging stages of this task is the construction of appropriate training samples which enable learning based on a limited number spectroscopically confirmed objects.
   } %leave it empty if necessary  
   {We describe how the \fink\  broker early supernova Ia classifier optimizes its ML classifications by employing an active learning (AL) strategy. We demonstrate the feasibility of implementation of such strategies in the current Zwicky Transient Facility (ZTF) public alert data stream.
   }
  % methods heading (mandatory
   {We compare the performance of two AL strategies: uncertainty sampling  and random sampling. Our pipeline consists of  3 stages: feature extraction, classification and learning strategy. Starting from an initial sample of 10 alerts (5 supernovae Ia, SNe Ia, and 5 non-Ia), we let the algorithm identify which alert should be added to the training sample. The system is allowed to evolve through 300 iterations.}
   {Our data set consists of 23 840 alerts from the ZTF with confirmed classification via cross-match with SIMBAD database and the Transient name server (TNS), 1 600 of which were SNe Ia (1 021 unique objects). The data configuration, after the learning cycle was completed, consists of 310 alerts for training and 23 530 for testing. Averaging over 100 realizations, the classifier achieved  $\sim$89\% purity and  $\sim$54\% efficiency. From 01/November/2020 to 31/October/2021 \fink\  has applied its early supernova Ia module to the ZTF stream and communicated promising SN Ia candidates to the TNS. From the 535 spectroscopically classified  \fink\ candidates, 459 (86\%) were proven to be SNe Ia. 
   }
   {Our results confirm the effectiveness of active learning strategies for guiding the construction of optimal training samples for astronomical classifiers. It demonstrates in real data that the performance of learning algorithms can be highly improved without the need of extra computational resources or overwhelmingly large training samples.  This is, to our knowledge, the first application of AL to real alerts data.}


   \maketitle
   
%
%-------------------------------------------------------------------

\section{Introduction}

Once large scale photometric astronomical surveys successfully overcome the challenge of data acquisition, the next bottleneck in the path for scientific exploration is to identify interesting candidates. This is specially important for researchers working on transient astronomy, where there is a window of opportunity when a potentially interesting target can be spectroscopically followed before becoming too dim to allow useful scrutiny. Depending on the specific astronomical target under investigation, this window can be as short as a few minutes.

Current surveys, like the Zwicky Transient Facility\footnote{\url{https://www.ztf.caltech.edu/}} (ZTF), publicly advertise detected candidates through an alert data stream distributed to community brokers, whose task is to filter/process and re-distribute interesting candidates with added values to the astronomical community. A similar protocol will also be adopted by the upcoming Vera C. Rubin Observatory Legacy Survey of Space and Time\footnote{\url{https://www.lsst.org/}} (LSST). However, due to the large data volume expected from LSST only 7 full copies of the stream will be distributed. The recently selected LSST brokers are {\sc ALeRCE} \citep{Forster:2020}, {\sc Ampel} \citep{ampel}, {\sc Babamul}, {\sc Antares} \citep{antares}, \fink\footnote{\url{https://fink-broker.org}}  \citep{fink}, {\sc Lasair} \citep{Smith_2019} and  {\sc Pitt-Google}. 

The \fink\ broker is being developed by an international community of researchers with a large variety of scientific interests, including among others multi-messenger astronomy, supernovae, solar system, anomalies identification, micro-lensing and gamma-ray bursts optical counterparts. It is currently processing and classifying alerts from ZTF as a test-bed for Rubin operations. In this work, we give a detailed description of the early supernova Ia classifier currently operating within \fink. 

The module uses an active learning strategy (AL) to select an optimum set of alerts that is subsequently used as training sample for a random forest classifier. The basic idea of AL is to allow the algorithm to identify which alerts should be labeled,  and included in the training set, according to their potential in improving the classifier. This philosophy is in the core of many of the recommendation systems currently in place, from pharmaceutical studies \citep{spjuth2021} to urban planning \citep{abernethy2018}.  In astronomy, it has proven to be effective in a few different scenarios \citep[e.g. ][]{solorio2005, richards2012,vilalta17,Walmsley2020}, including the classification of astronomical transients \citep[e.g. ][]{ActSNClass-paper, kennamer2020} and anomaly detection \citep{ishida2021}.

The remaining sections of this paper are organized as follows: Section \ref{sec:data} describes the construction of our data set. Section \ref{section:Methodology} describes all stages of our methodology, including filtering (\ref{sec:filter}), feature extraction (\ref{sec:features}), classifier (\ref{sec:class}), learning strategy (\ref{sec:al}) and evaluation metrics (\ref{sec:metrics}). We show our results in Section (\ref{sec:results}) and discuss their implications and future plans for \fink\ in Section \ref{sec:conclusions}.

\begin{figure}
\noindent \begin{minipage}[t]{\columnwidth}
\includegraphics[width=\textwidth]{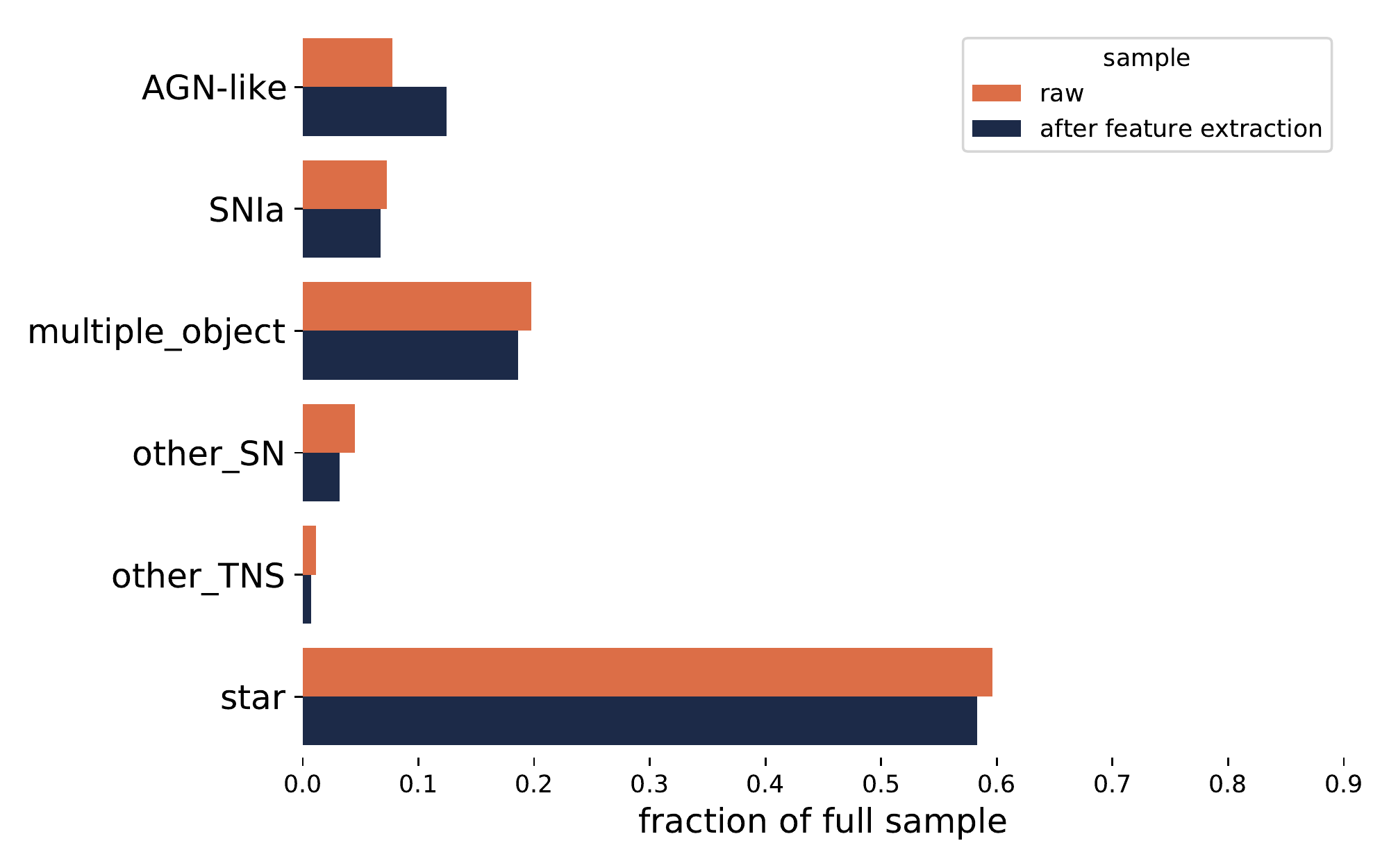}
\captionof{figure}{Fraction of each class within the original data set before processing (orange) and the sample surviving feature extraction (blue). Numerical values are shown in Table \ref{tab:pop}.} 
\label{fig:pop_features}
\vspace{2ex}
\begin{tabular}{c|c|c}
       Class & in raw data & after feature extraction \Bstrut \\
       \hline
        \texttt{AGN-like}         & 6 677 (7.7)   & 2 961 (12.5) \Tstrut\\
        \texttt{SNIa}             & 6 269 (7.3)   & 1 600 (6.7) \\
        \texttt{other\_SN}        & 3 891 (4.5)   & 762 (3.2) \\
        multiple\_object & 17 083 (19.8)  & 4 419 (18.6)\\ 
        \texttt{other\_TNS}       & 964 (1.1)     & 174 (0.7) \\
        star             & 51 538 (59.6)  & 13 859 (58.3) \Bstrut\\
        \hline
        Total            & 86 422 & 23 775 \Tstrut
    \end{tabular}
    \vspace{1ex}
    \captionof{table}{Population types in our alerts data set before and after feature extraction. Values in parenthesis indicate percentage of total sample. Note that the same object on the sky can emit several alerts. As a reference, the total number of objects in the raw data (after feature extraction) is \referee{52 405 (15 751)}, while the total number of objects for \texttt{SNIa} in the raw data (after feature extraction) is 2 287 (1 021).}
    \label{tab:pop}
\end{minipage}
\end{figure}

%--------------------------------------------------------------------
\section{Data}
\label{sec:data}

\referee{Our starting point was a data set containing all public alerts from the ZTF survey from November/2019 to March/2020 with confirmed classifications available through the SIMBAD catalog \citep{SIMBAD:2000} or Transient Name Server\footnote{\url{https://www.wis-tns.org/}} (\referee{TNS}) database. \review{TNS classified objects are spectroscopically confirmed while SIMBAD contains highly curated labels from both spectroscopic and photometric surveys \citep{Loup:2019}, with the majority of its classifications taking into account spectroscopic redshift measurements (CDS, private communication)}. 
\review{Each alert contains the last 30 days history of photometric measurements for a given point in the sky}. %\am{which include limiting magnitudes}. 
From the SIMBAD data set, this amounts to 5 004 378 alerts corresponding to 1 313 371\footnote{The same object can emit several alerts over time, and each alert carries the photometric history over the last 30 days, thus allowing light curve analysis.} objects, with dominant classes being RRLyr ($\approx$26\%), eclipsing binaries ($\approx$22\%) and stars ($\approx$19\%)). From the TNS data, there were 11 422 alerts corresponding to 11 329 objects, dominated by SNIa (~52\%) and SNII (19\%)). }

\referee{From the TNS set, all alerts were used. Our experiments showed  that including around 5 or 6 times the same number of alerts with SIMBAD classification was enough to construct an informative training sample. Using a larger number of SIMBAD sources does not affect the main results of a binary classification study as the one presented here. In order to be conservative while not overusing computer resources, we randomly selected 75 000 alerts, corresponding to 41 076 objects}\footnote{\referee{The number of SIMBAD objects to be used is a parameter input in our code and can be easily changed to allow further scrutiny, e.g. for users interested in multi-class experiments.}} In what follows, this combined set will be referred to as raw data. 

We used the photometric information contained in alerts, regarding ZTF broad-band filters [$g,r$], retrieved through the \fink\  API\footnote{\url{https://fink-portal.org/api}}. According to the \fink\  schema, these correspond to alert id (\texttt{candid}), time of observation in Julian dates (\texttt{jd}), magnitude from PSF-fit photometry (\texttt{magpsf}), error in magnitude (\texttt{sigmapsf}) and ZTF filter  (\texttt{fid}) as well as their TNS (\texttt{TNS}) and SIMBAD (\texttt{cdsxmatch}) associated classes. \review{In order to avoid classification errors between the Ia and non-Ia classes, we did not use SN from Simbad.} \review{Thus all SN used in this work are spectroscopically classified objects from TNS.}
% Thus rendering all SN being spectroscopically confirmed}. 
The full raw data set consisted of 86 422 alerts, corresponding to 52 405 unique sources. 

Table \ref{tab:pop} shows the population of classes within this data set. Alerts possessing a SIMBAD cross-match were divided into 3 categories (\texttt{star}, \texttt{multiple\_objects}\footnote{Multiple objects include eclipsing binaries, cataclysmic variables and novae.} and \texttt{AGN-like} following SIMBAD's classification scheme\footnote{\url{http://simbad.u-strasbg.fr/simbad/sim-display?data=otypes}}$^{,}$\footnote{We use the term \texttt{AGN-like} for all cross-matches related to AGNs: QSO, Blazars, etc.}. Those associated with supernovae (not Ia) were allocated in the \texttt{other\_SN} category while all SN Ia are reported under the \texttt{SNIa} class (it comprises   \texttt{SN Ia, SN Ia-91T-like, SN Ia-91bg-like, SN Ia-CSM, SN Ia-pec},  \texttt{SN Iax}/\texttt{02cx-like}). Additional transients whose classification was provided via TNS were grouped under the \texttt{other\_TNS} class including Cataclismic Variable Stars, TDEs, etc. We made this raw data set publicly available, along with the necessary code to reproduce the results in this paper,  through Zenodo\footnote{\url{https://doi.org/10.5281/zenodo.5645609}}.

\begin{figure*}[h!]
    \centering
    \subfigure{\includegraphics[width=0.495\textwidth]{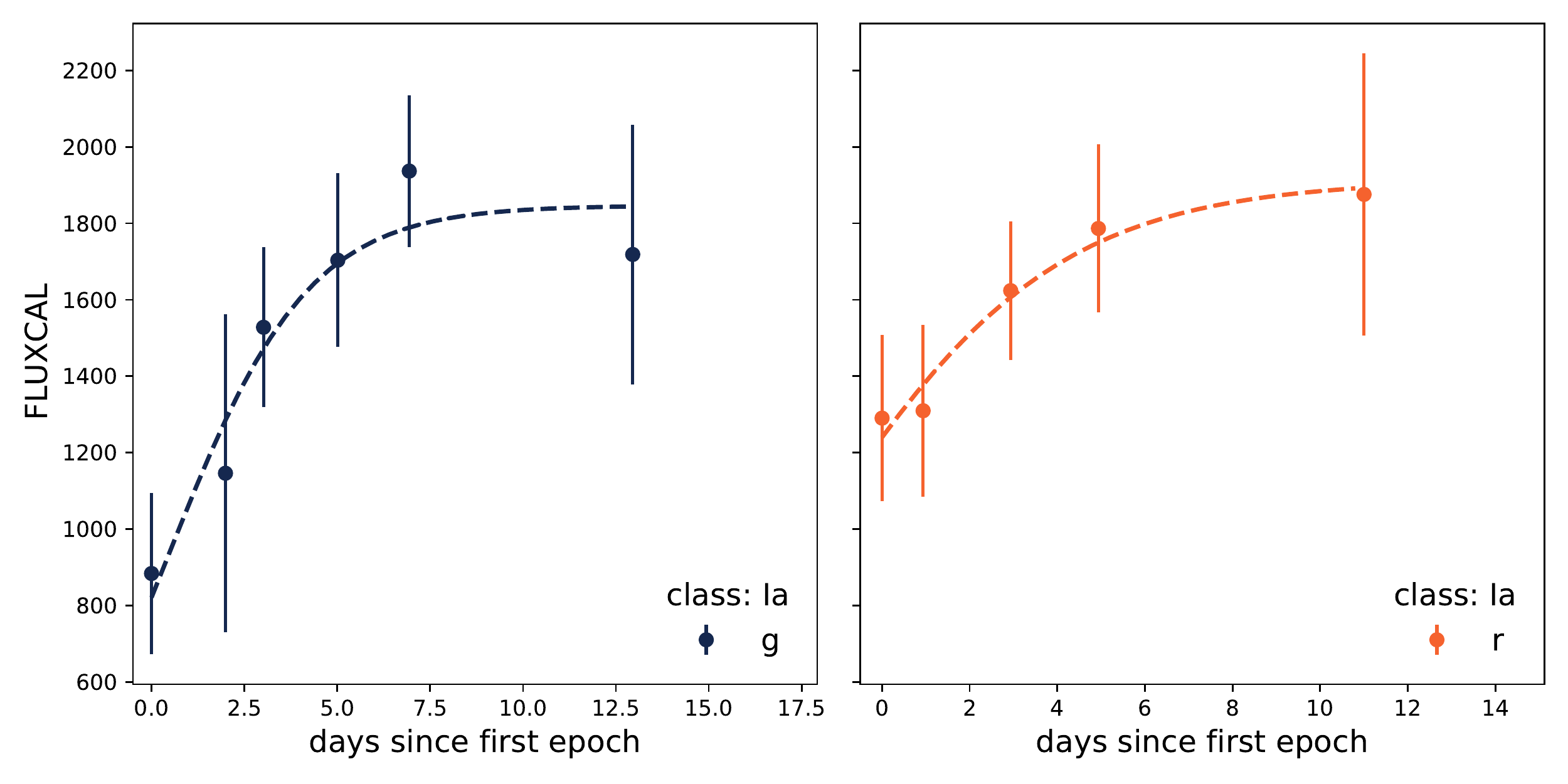}}
    \subfigure{\includegraphics[width=0.495\textwidth]{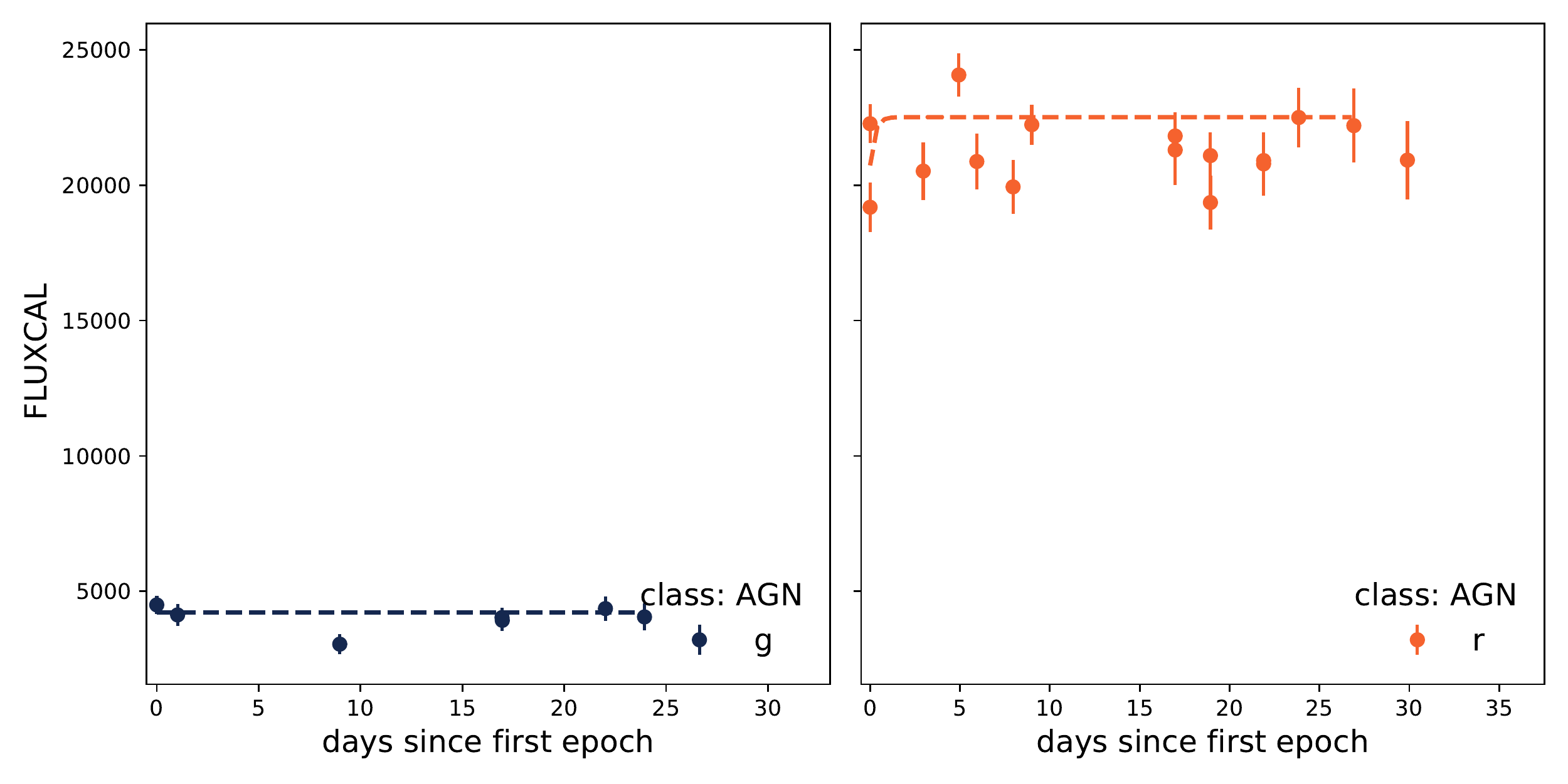}}
    \subfigure{\includegraphics[width=0.495\textwidth]{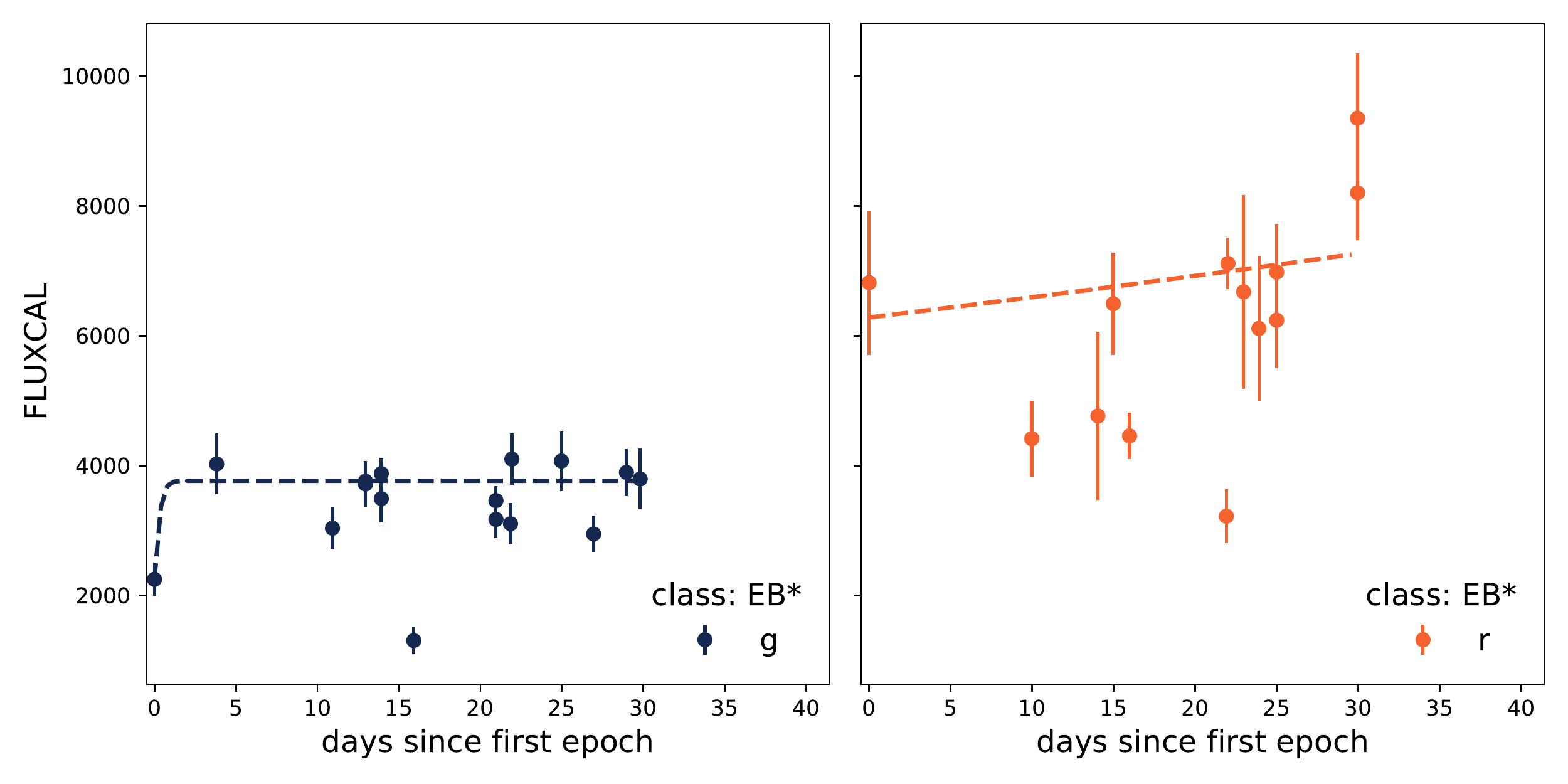}}
    \subfigure{\includegraphics[width=0.495\textwidth]{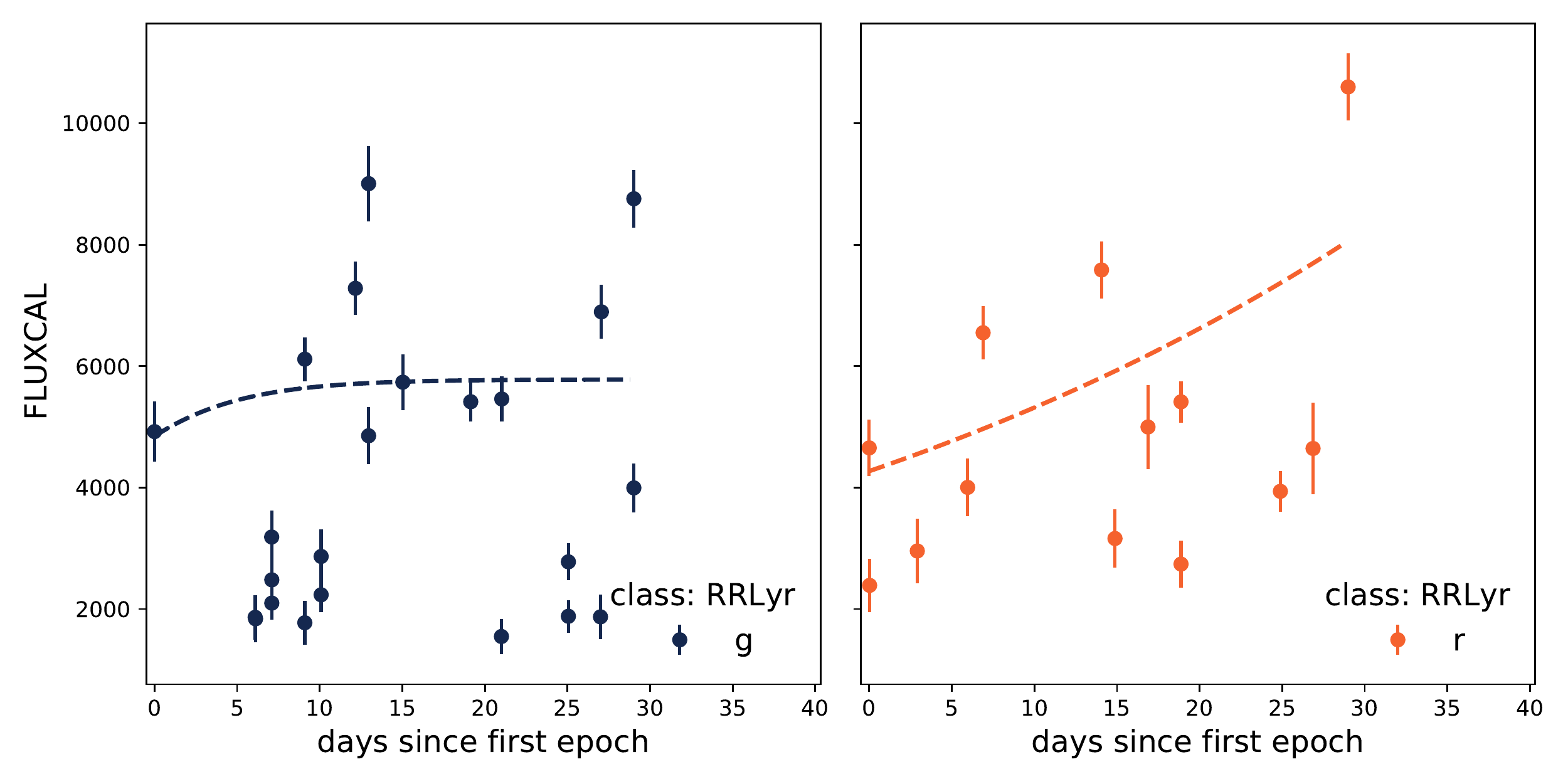}}
    \caption{Alert data and sigmoid  fit for alerts associated with 
    SNe Ia (\texttt{objectId}:\referee{ZTF20acqjmzk}, \texttt{candid}: \referee{1424192755315015009}), 
    Galaxy (\texttt{AGN-like}) (\texttt{objectId}:\referee{ZTF18aarybyq}, \texttt{candid}: \referee{1452439215115010000}), 
    eclipsing binary (\texttt{objectId}:\referee{ZTF18aabtvle}, \texttt{candid}: \referee{1277247161915010000}) and RRLyr star (\texttt{objectId}:\referee{ZTF18aaptrep}, \texttt{candid}: \referee{1385238050715015010}). Points and error bars denote the observed values. Dashed lines show approximations obtained from the sigmoid fit (Section \ref{sec:features}). \referee{The corresponding best-fit parameters are given in Appendix \ref{app:features}.} }
    \label{fig:fit}
\end{figure*}

\section{Methodology}
\label{section:Methodology}

We \review{adapted} the active learning approach presented in \citet{ActSNClass-paper} %\am{eliminate: in order} 
to optimize the construction of an effective training sample. Starting from their basic template pipeline of feature extraction, classification and learning strategies, we performed a few adaptations to accommodate the higher complexity intrinsic to real data sets. 

Since we are mostly interested in early classification (identifying SN Ia before they reach maximum brightness), we begin by \review{selecting only} 
% filtering the 
epochs corresponding to the rising part of the light curve (Section \ref{sec:filter}). Subsequently, we developed a feature extraction method based on a sigmoid function (Section \ref{sec:features}) which requires \review{at least} 3 observed epochs per filter -- 
\review{less than the minimum 5 observations needed to fit the parametric function used in}
% less demanding when compared to the 5 parameter fit used in
\citet{ActSNClass-paper}. The numbers, and percentages, of each sub-class surviving this preprocessing are shown in Table \ref{tab:pop} and Figure \ref{fig:pop_features}. 

\begin{figure*}[h!]
    \centering
    \includegraphics[width=\textwidth]{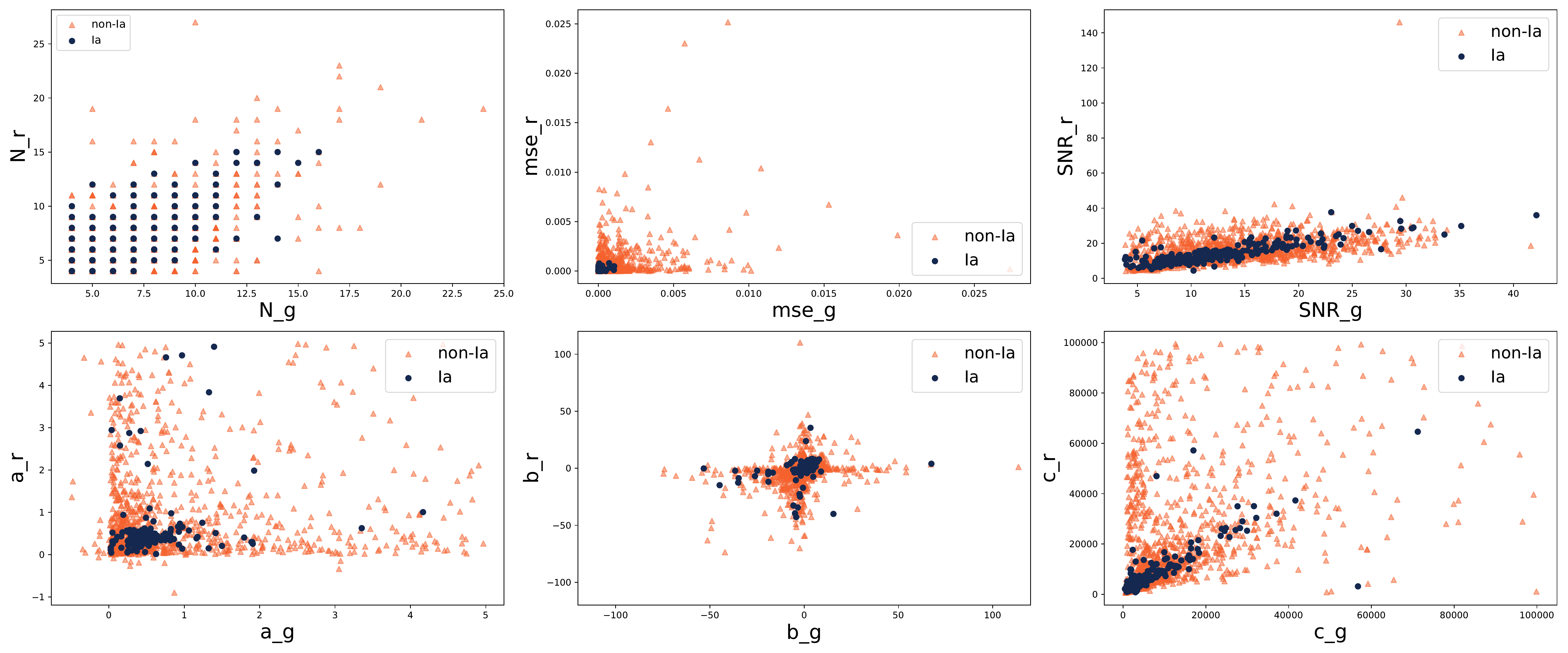}
    \caption{Distribution in %%\am{two-dimensional} 
    feature space according to alert true classes. Each panel show the same parameter in {\tt g} and {\tt r} bands (horizontal and vertical axis, respectively). The blue dots  represent alerts from SN Ia sources, while the orange triangles represent all other classes in our initial data set. We can observe that non-Ias occupy an overlapping and larger region of the parameter space than its Ia counterparts.}
    \label{fig:features}
\end{figure*}

\subsection{Filtering}
\label{sec:filter}

In order to isolate the alerts depicting light curves on the rise we treat each filter separately. An exponentially weighted moving average, with a window of \referee{4} points was performed, and its derivative \review{(i.e. forward finite difference)} calculated. Epochs corresponding to a negative derivative were discarded. 
\review{This process ensures that at least 3 points are in the rise, 
as one point will serve as the comparison baseline for the  finite difference. 
}

This procedure is  successful in masking out the declining part of the light curve for transient events, however it can fail to grasp details of periodic or oscillatory light curves (examples of outcomes from this filtering process are shown in Figure \ref{fig:fit} and the corresponding best-fit parameter values are shown in Appendix \ref{app:features}). Nevertheless, it proved to be successful in filtering enough information to allow supernova-like events to populate a specific area of the parameter space when combined with the feature extraction procedure described in Section \ref{sec:features}.

\subsection{Feature extraction}
\label{sec:features}

All surviving epochs were converted from ZTF magnitudes $m$ and magnitude error $\Delta m$ into SNANA\footnote{\url{https://github.com/RickKessler/SNANA/blob/master/doc/snana_manual.pdf}} \citep{snana} flux units,  \review{$f$ and its corresponding error $\Delta f$ derived by the chain rule},
\begin{eqnarray}
    f & = & 10^{-\frac{4m}{10} + 11}, \\
    df & =& \frac{4 \ln{(10)}}{10}   * dm * f \nonumber \\
     & = & 10 ^{10} * \alpha * dm * \exp{\left(-\frac{\alpha m}{10} \right)}
    \label{eq:fluxcal}
\end{eqnarray}
with ${\alpha = 4 \times \ln{10} \sim} 9.21034$. Subsequently, observations in each filter were independently submitted to a sigmoid fit \review{$S$ at a given time $t_i$}, 
\begin{equation}
{S}(t_i) = \dfrac{c}{1 + e^{-a (\Delta{t}_i - b) }},
\label{eq:fit-sigmoid}
\end{equation}
where $\Delta t_i = t_i - \min(t)$  is the observation time of the $i$-th data point since first detection. Employing a least square minimization routine and initialization values \review{$a_0,b_0,c_0$ defined by}
\begin{eqnarray}
 a_{0} & = &  \frac{\max(f) - \min(f)}{{t_{f_{\rm max}}-\min(t)}}, \\
b_{0} & = & \frac{1}{a_{\rm 0}}\log\left[{\frac{c_{\rm 0}}{min(f) - 1}}\right], \\
 c_{0} & = & \max(f), 
 \end{eqnarray}
where $N$ is the total number of points surviving the filtering for this alert \review{and $t_{f_{max}}$ is the time of maximum flux for a given light-curve}, we obtained the 3 features for each alert and band from the best fit values of $a,b$ and $c$. Three other features were extracted per band:

\begin{enumerate}[i)]
\item the quality of the fit, represented by
\begin{equation}
    {{\rm MSE} = \sum_{i=1}^N \frac{(\tilde{S}_i - \tilde{f}_i)^2}{\tilde{S}_i}},
\end{equation}
where $S_i$ is the flux estimate for the $i$-th epoch using the best fit values for $a,b$ and $c$, and tilde quantities 
% are related to underlined ones by
\review{( $\tilde{S}_i$ and $\tilde{f}_i$) generalized as $\tilde{X}_i$ are normalized values of $X_i$ defined by}
\begin{equation}
    \tilde{X}_i = \dfrac{X_i}{\sum_j X_j};
\end{equation}

\item the mean signal to noise ratio (SNR), 
\begin{equation}
    {\rm \texttt{SNR}} = \frac{1}{N}\sum_{i=1}^N \frac{f_i}{\Delta f_i}
\end{equation}
\item the total number of epochs used in the fit, $N$. 
\end{enumerate}
   
In summary, for each alert we have a total of 6 parameters, $[a,b,c,\chi^2,$\texttt{SNR}$,N]$, per band. The input matrix is constructed by concatenating the parameters corresponding to $[g,r]$ bands for the same alert in one line. 

\review{Figure \ref{fig:features} shows the  two-dimensional representation of objects in this parameter space.} Our final input matrix is composed of \referee{23 775} lines (alerts) and 12 columns (features), this corresponds to \referee{15 751} unique sky objects. The composition of this alert sample in classes is shown in Table \ref{tab:pop}. From this table, we noticed that although the feature extraction reduced considerably the data volume, it maintained the overall proportion between classes present in the original raw data.

\subsection{Classifier}
\label{sec:class}

Following the framework outlined in \citet{ActSNClass-paper}, we used a Random Forest classifier \citep{Breiman01}. This is an ensemble method which uses a number of decision trees \citep{breiman1984}, constructed from different sub-samples of the training set. Once the forest is constructed, the estimated classes for the target sample are determined by majority voting considering all trees in the ensemble. The method proved to be effective in a variety of data scenarios in astronomy \citep[e.g., ][]{richards2012,ishida2013,lochner2016,Moller2016,Calderon2019,Nixon2020,Kuhn2021,Nakazono2021}. 

Random Forest has a number of advantages when faced with complex data sets. Since it is based on decision trees, the trained model is fully interpretable, which allows each decision to be scrutinized. Moreover, for the specific purpose of this project, its most important quality is the sensitivity to small changes in the training set. Decision trees divide the parameter space into small regions around each object in the training sample. Thus, in the small training data regime, it quickly adapts classification results when faced with a small number of new labels. This is a crucial feature for any classifier which needs to work within an active learning framework (Section \ref{sec:al}). All Random Forest models trained in this work were constructed using 1000 trees, to ensure good convergence in a reasonable time.

\begin{figure*}
    \centering
    \noindent \begin{minipage}[t]{\textwidth}
    \centering
    \includegraphics[width=0.75\textwidth]{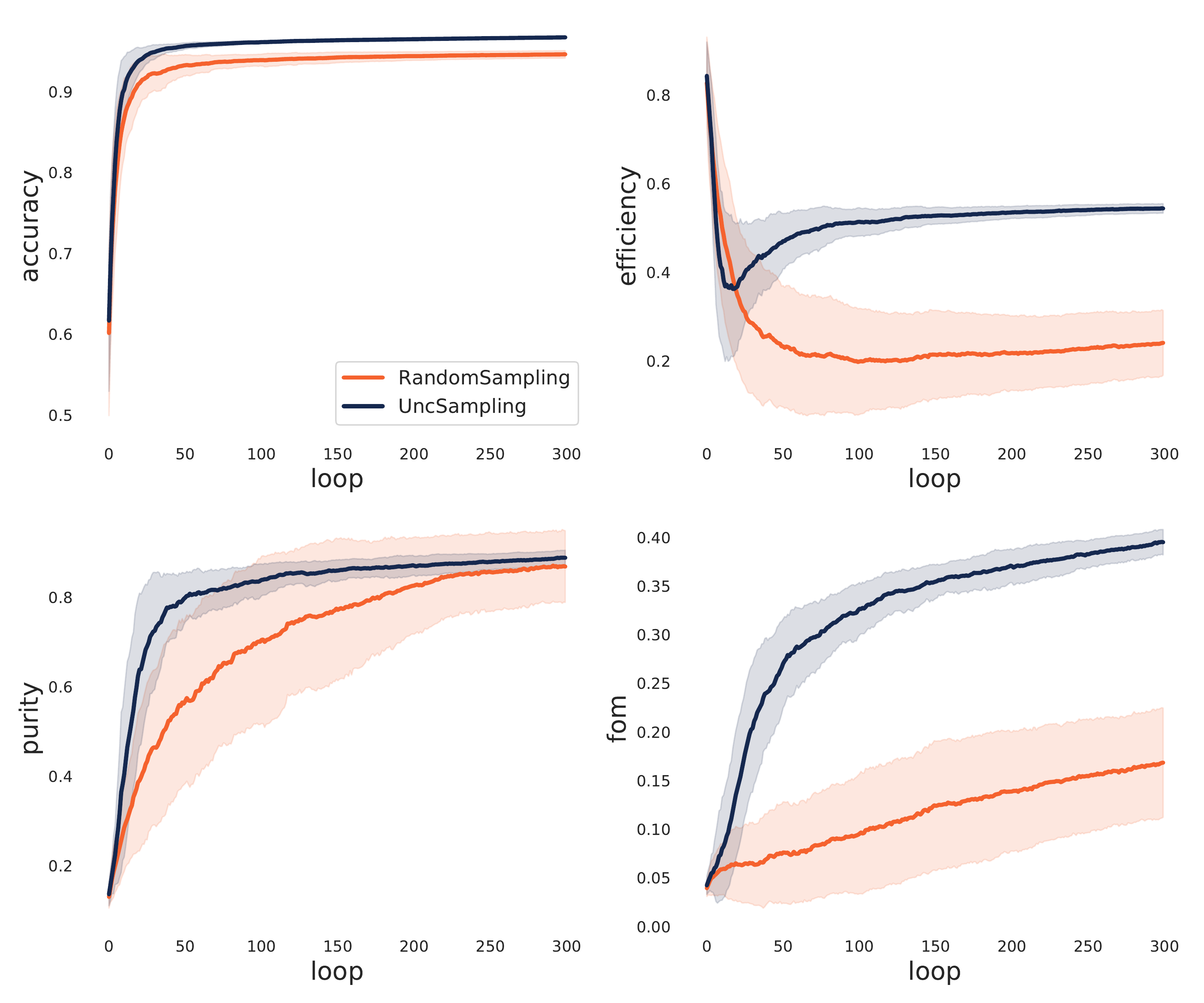}
    \captionof{figure}{Evolution of classification results as a function of the active learning loop. In all panels the orange and blue lines show results from random and uncertainty sampling, respectively. The dashed regions mark 1 standard deviation over 100 realizations. The initial training sample in all experiments consisted of 10 alerts (5 randomly selected from each class) and at each loop only 1 object was added to the training sample (batch = 1).}
    \label{fig:metrics}
    \end{minipage}
    \begin{minipage}{\columnwidth}
    \centering
    \bigskip
    \begin{tabular}{c|c|c}
              &  \multicolumn{2}{c}{Learning Strategy} \Bstrut\\
        Metric & Random sampling & Uncertainty sampling \Bstrut \Tstrut\\
        \hline
        accuracy   & 0.95 $\pm$ 0.01 &  0.97 $\pm$ 0.01  \Bstrut \Tstrut \\
        efficiency & 0.24 $\pm$  0.07 & 0.54 $\pm$ 0.01 \Bstrut \Tstrut \\
        purity     & 0.87 $\pm$ 0.08 &  0.89 $\pm$ 0.02 \Bstrut \Tstrut \\
        fom        & 0.17 $\pm$ 0.06 & 0.39 $\pm$ 0.01\Bstrut \Tstrut  \\
    \end{tabular}
    \bigskip
    \captionof{table}{Mean and standard deviation results for the 2 query  strategies investigated in this work. Reported values correspond to the final state of the model, after 300 iterations, averaged over 100 realizations. A full representation of the classification results in intermediate steps is given in Figure \ref{fig:metrics}.}
    \label{tab:metric}
    \end{minipage}
    \begin{minipage}{\columnwidth}
    \centering
    \includegraphics[width=\columnwidth]{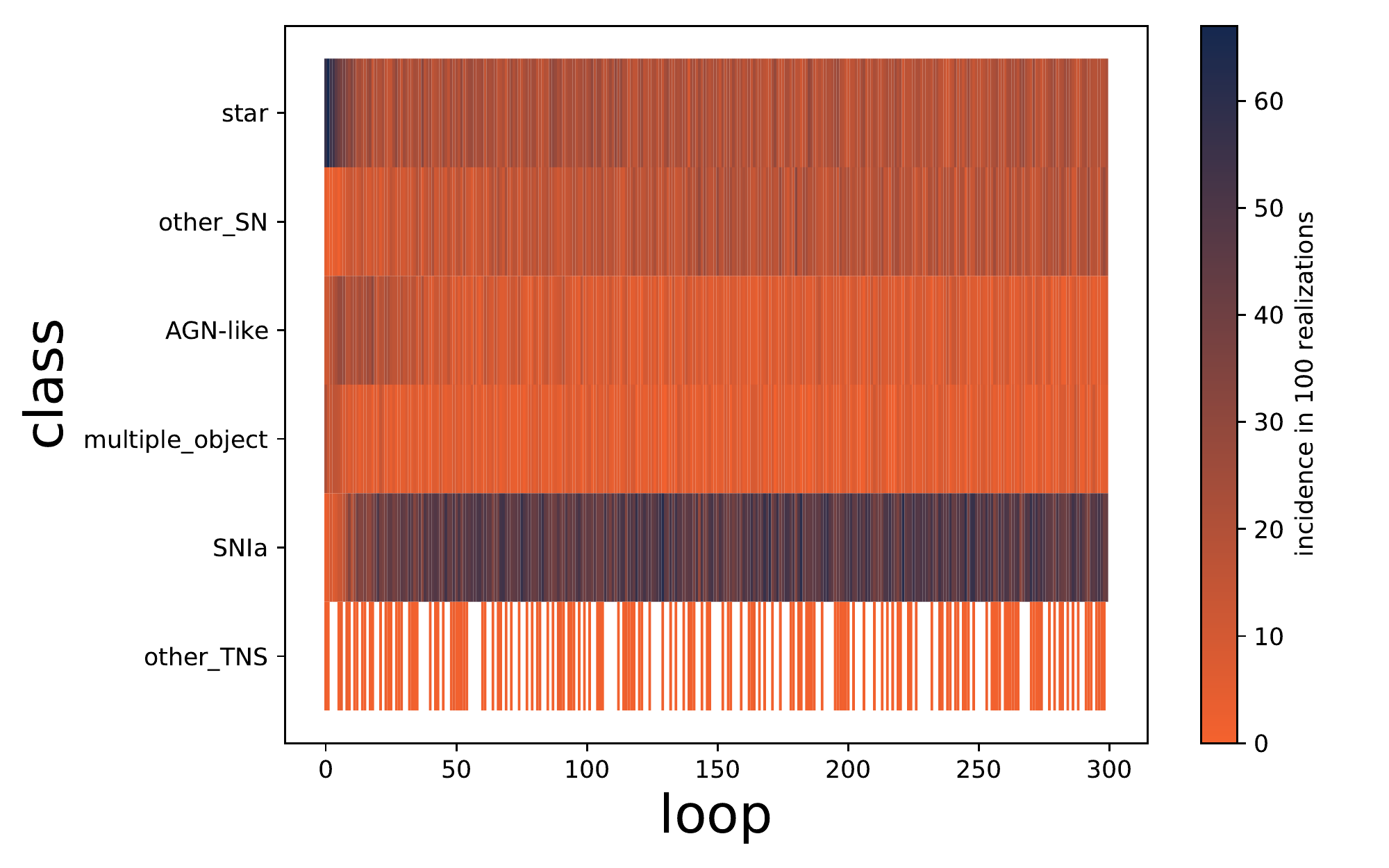}
    \captionof{figure}{Classes of the queried alerts as a function of the learning loop. The plot shows queries selected by the uncertainty sampling strategy over 100 realizations.}
    \label{fig:classes_loop}
    \end{minipage}
\end{figure*}
\subsection{Active Learning}
\label{sec:al}

Active learning \citep[AL, ][]{Settles12} is a branch of machine learning methods whose goal is to optimize classification results when labeling is expensive and/or time consuming. In its pool-based approach, at each iteration the algorithm identifies objects which can potentially increase the information content available in the training sample if labels were provided. This list is sent to an annotator (which can be a human or a machine) who will be in charge of labeling the \textit{queried sample}. Once labels are available, the queried sample is added to the training, the machine learning model is re-trained and the process is repeated. The iterations continue until the labeling resources (\textit{budget}) are exhausted. 

In the context of astronomical transient classifications, we start with a small number of \review{classified transients in available catalogs.} Subsequently, the algorithm will identify among all non-labeled objects which of them would be more informative to the model if a classification (label) was provided. Once this queried object is identified and \review{its classification} 
% its spectroscopic confirmation 
is obtained, we %return to the machine learning model, 
add the new photometry and label to the training sample and re-train the \review{learning} model. Each iteration in this process is also called a "loop". \review{We highlight that, although here this exercise is performed in catalog data, our intend during LSST production is to obtain labels through spectroscopic follow-up.} %if above is true, then a good way to clarify what is what for the referee

AL strategies differ on how the most informative objects are chosen (\textit{query strategy}). In this work, we used  uncertainty sampling \citep[US, ][]{sharma2017}, which identifies the objects closest to the decision boundary between classes at each iteration. We are focused on correctly identifying SNe Ia, thus, our experiment is designed as a binary classification  problem (Ia vs non-Ia). At each iteration the algorithm will query the object with highest entropy \citep[for a binary classifier, this correspond to the object with classification probability closest to 0.5. See, e.g., ][]{MacKay2003}.

We also perform experiments using a random sampling \referee{(RS)} approach, where at each iteration objects are randomly selected from the target and added to the training sample. This allow us to better understand how much influence the decision process has in the final classification results.

\subsection{Metrics}
\label{sec:metrics}

We evaluate our results using the classification metrics and nomenclature from the Supernova Photometric Classification Challenge \citep{kessler2010},
\begin{eqnarray}
{\rm accuracy} & = & \frac{C_{\rm Ia} + C_{\rm non-Ia}}{N}, \\
{\rm efficiency} & = & \frac{C_{\rm Ia}}{N_{\rm Ia}}, \\
{\rm purity} & = & \frac{C_{\rm Ia}}{C_{\rm Ia} + W_{\rm non-Ia}} \qquad {\rm and}\\
{\rm fom} & = & {\rm efficiency} \times \frac{C_{\rm Ia}}{C_{\rm Ia} + W^{*} \times W_{\rm non-Ia}},
\end{eqnarray}
where \texttt{fom} stands for figure of merit, $C_{\rm Ia}$ is the number of correctly classified Ias, $C_{\rm non-Ia}$ denotes the number of correctly classified non-Ias, $N$ represents the total number of objects in the target sample, $N_{\rm Ia}$ is the total number of Ias in the target sample, $W_{\rm non-Ia}$ is the number of wrongly classifier non-Ias and $W^*=3$ \citep{kessler2010} is a weight which penalizes false positives. 

We stress that in our work these metrics are used for diagnostic purposes only and are not used in the decision making process. 

\begin{table}
\centering
\begin{tabular}{c|c|c}
              & \multicolumn{2}{c}{Learning Strategy}  \\
    \multirow{2}{*}{Class}  & Random  & Uncertainty \Tstrut \\
    & sampling &  sampling \Bstrut \\
    \hline
    \texttt{SNIa}               &  6.58 $\pm$ 1.28 &  44.80 $\pm$  1.40  \Tstrut \Bstrut \\
    \texttt{star}               &  58.52 $\pm$ 2.89 & 21.58 $\pm$  2.04  \Tstrut \Bstrut \\
    \texttt{other\_SN}          &  3.17 $\pm$ 0.93 & 16.76 $\pm$ 1.70  \Tstrut \Bstrut \\
    \texttt{multiple\_object}   & 18.84  $\pm$ 2.35 & 6.31 $\pm$ 1.28   \Tstrut \Bstrut \\
    \texttt{AGN-like}           &  12.15 $\pm$ 1.95 & 9.82 $\pm$ 1.95  \Tstrut \Bstrut \\
    \texttt{other\_TNS}         &  0.82 $\pm$ 0.46 & 0.78 $\pm$  0.36 \Tstrut \Bstrut \\
    \end{tabular}
    \bigskip
     \caption{Percentages of each class in the final queried sample. Values show mean and standard deviation over 100 realizations.}
    \label{tab:queried}
\end{table}

\begin{figure}
\noindent \begin{minipage}[t]{\columnwidth}
\includegraphics[width=\textwidth]{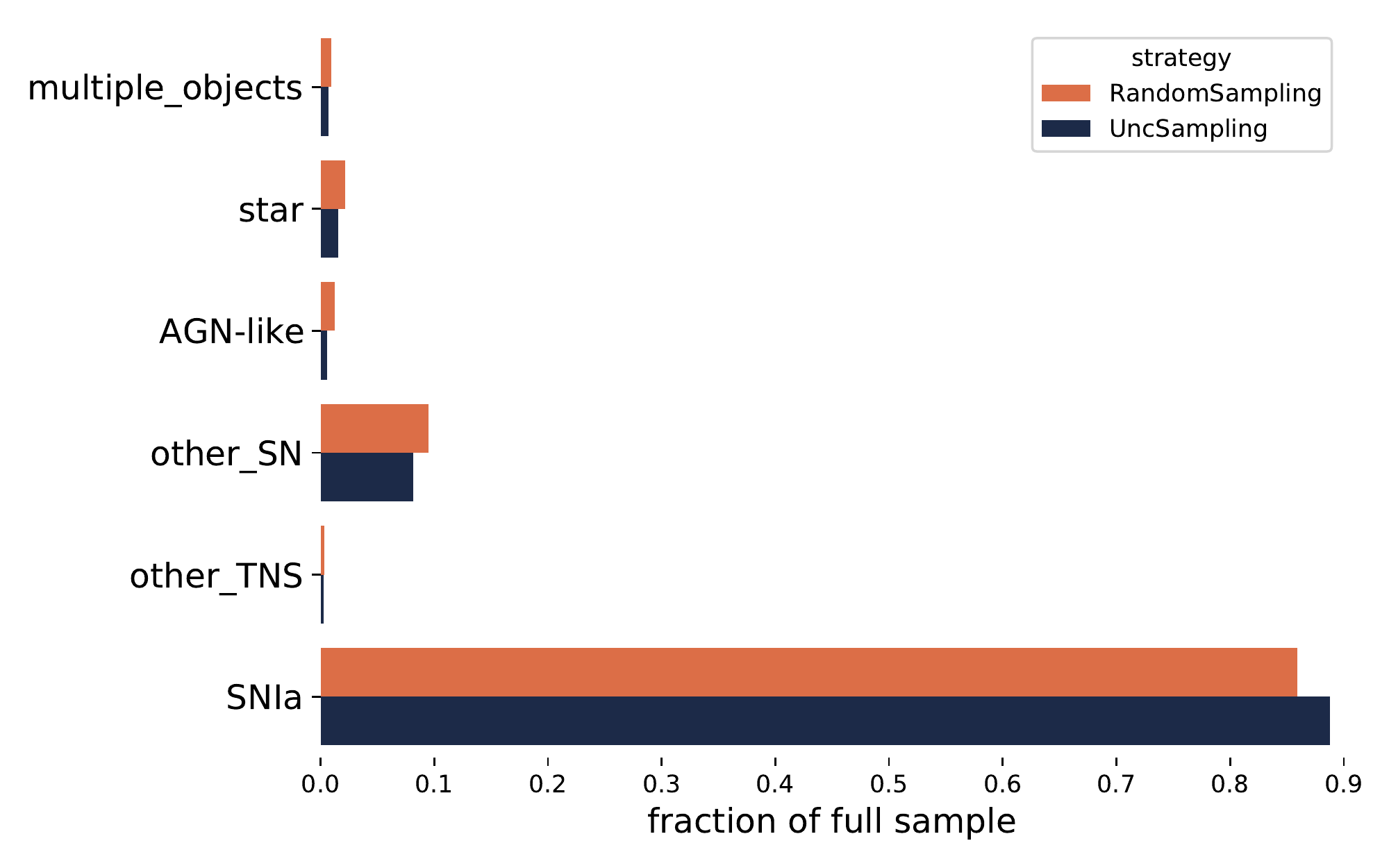}
\captionof{figure}{Distribution of true classes among the alerts classified as SN Ia. The classifier was trained using the complete queried in addition to the initial training sample (after 300 learning loops) gathered by the random (orange) and uncertainty (blue) sampling strategies. The bar length shows mean over 100 realizations. Numerical values are given in Table  \ref{tab:contamination}.}
\label{fig:photoclass}
\vspace{2ex}
\centering
\begin{tabular}{c|c|c}
              & \multicolumn{2}{c}{Learning Strategy}  \\
    \multirow{2}{*}{True Class}  & Random & Uncertainty  \Tstrut \\
    & sampling & sampling \Bstrut\\
    \hline
    \texttt{multiple\_object}  & 0.95 $\pm$ 0.66 & 0.66 $\pm$ 0.29  \Tstrut \Bstrut\\
    \texttt{star}              & 2.13 $\pm$ 1.29 &  1.51 $\pm$ 0.39  \Tstrut \Bstrut\\
    \texttt{AGN-like}          & 1.25 $\pm$  1.39 & 0.53 $\pm$ 0.24 \Tstrut \Bstrut\\
    \texttt{other\_SN}         & 9.50 $\pm$ 1.73 & 8.25 $\pm$  0.74 \Tstrut \Bstrut\\
    \texttt{other\_TNS}        & 0.35 $\pm$ 0.10 & 0.25 $\pm$  0.09 \Tstrut \Bstrut\\
    \texttt{SNIa}              & 85.99 $\pm$ 3.89 & 88.80 $\pm$ 1.07 \Tstrut \Bstrut\\
\end{tabular}
\bigskip
\captionof{table}{Percentages of each true class in the final set of alerts photometrically classified as SNe Ia. Values show mean and standard deviations over 100 realizations.}
\label{tab:contamination}
\end{minipage}
\end{figure}

\section{Results}
\label{sec:results}

Starting from the initial data matrix described in Section \ref{section:Methodology}, we repeated the same experiment 100 times. In each realization, we randomly selected 10 objects (5 SNe Ia and 5 non-Ia), to be used as the initial training sample, and evolve the system through 300 iterations. The final configuration consists of 310 alerts in the training sample and \referee{23 425} in the test sample. This small number of iterations was enough to significantly improve the classification results while, at the same time, ensure that overall statistical properties of the target sample were maintained throughout the learning process\footnote{In data sets where such properties cannot be sustained through the system evolution, one should use an experiment design with separated pool and validation samples, as shown in \citet{kennamer2020}.}. The experiments were performed following \referee{RS} and \referee{US} query strategies. Figure \ref{fig:metrics} shows the evolution of classification metrics as a function of the learning loop. It begins with a very high efficiency ($\sim 0.80$) and low purity ($\sim 0.05$). This is a consequence of the small and equally balanced training sample and the large and highly unbalanced test sample. In this situation, random forest returns a high incidence of false positives (non-Ia classified as Ia). Nevertheless, since the number of non-Ias is very high, the initial accuracy is $\geq 0.60$. As the system evolves and more information is added to the training sample, the model quickly improves. This first burning phase lasts for $\sim 50$ loops, with 1 object being added to the training set per loop. In this phase, results from the 2 strategies do not differ significantly (e.g., efficiency decreases in both cases). Once general characteristics are learned the model begins to grasp information about details of the class boundaries and results start to differ strongly (e.g., efficiency values start to improve for \referee{US} and purity values grow apart for different query strategies). \referee{The system finally comes to a quasi-asymptotic behavior with \referee{US} maintaining a much higher efficiency value while both strategies present similar purity levels. This leads to \referee{US} resulting in a figure of merit more than two times higher than the value achieve by RS (Table \ref{tab:metric}).}

\referee{Qualitatively, these finds are in agreement with the full light curve results presented in \citet[][ figure 6]{ActSNClass-paper}. However, we see a significant decrease in the effectiveness of the RS strategy when compare to the results archived by US. In \citet{ActSNClass-paper}, RS figure of merit results achieve between 70\% and 80\% the levels reported by US, while Table \ref{tab:metric} shows RS figures stabilizing in  approximately 40\% of US ones. Despite different feature extraction and data sets, this behavior is mostly driven by the composition of the target sample. The ideal binary classifier should be  trained in an equally balanced data set.  In \citet{ActSNClass-paper}, the sample available for query was composed of $22\%$ \review{SN Ia} (Figure 3 of that paper), while in our results they represent merely 7\% of the objects available for query (Table \ref{tab:pop}). Thus, in that work RS results in a training sample closer to the 50/50 fraction and consequently lead to better classification results.}

\referee{More} details about the queried sample are shown in Table \ref{tab:queried}. The RS strategy, as expected, queries individual classes following their fraction in the original data (Table \ref{tab:pop}). On the other hand, US queries \referee{$>7$} times more \texttt{SNIa} and more than 5 times as many the \texttt{other\_SN} type. Figure \ref{fig:classes_loop} shows how the queried classes change as a function of the learning loop. In early stages the queries are basically random, thus until loop 20 we see again an agreement with the overall class population (larger fraction of classes {\tt star} and {\tt AGN-like}). Once information is gathered about the larger populations, learning is concentrated in the class of interest, with larger fraction of SN Ia being queried in latter loops. 

Almost half of the alerts queried by US in each full learning cycle \referee{($\sim 135$)} were Ias. Thus, we can estimate that from the \referee{23 425} alerts in the test sample, Ias correspond to \referee{$\sim 1460$}. A detailed analysis of the photometrically classified set shows that, over 100 realizations, RS identifies \referee{$443 \pm 145$} alerts as Ias, while US photometrically assigns \referee{$896 \pm 33$} alerts to the \texttt{SNIa} class. Given that both strategies report similar purity values, this means an almost double increase in the number of SNIa correctly classified \referee{when the training sample is constructed}  using the US query strategy. 
Figure \ref{fig:photoclass} and Table \ref{tab:contamination} show the average composition of the final photometrically classified Ias. 

\begin{figure*}
    \centering
    \begin{minipage}{\textwidth}
    \centering
    \includegraphics[width=0.5\textwidth]{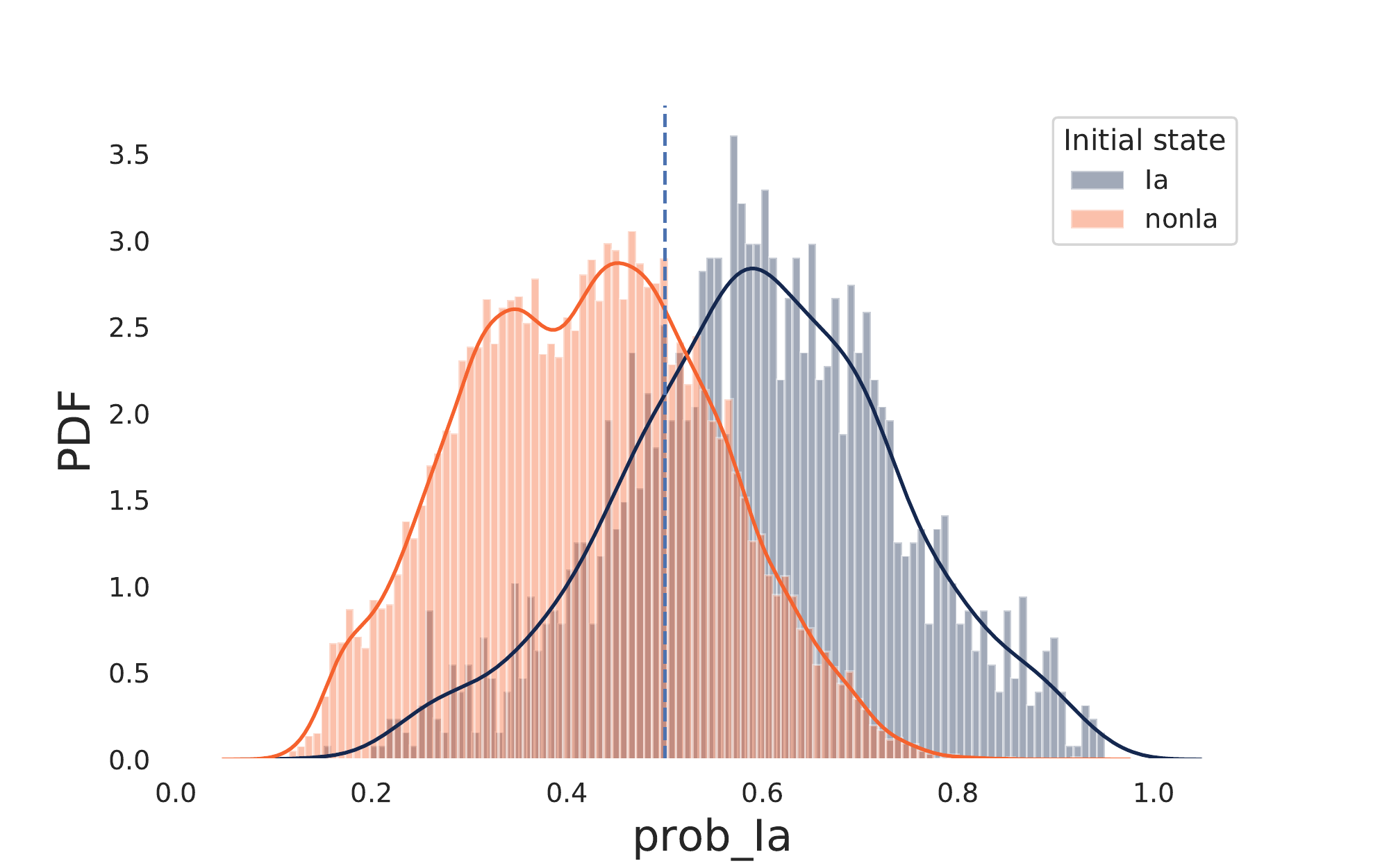}
    \end{minipage}
    \begin{minipage}{\textwidth}
    \centering
    \includegraphics[width=\textwidth]{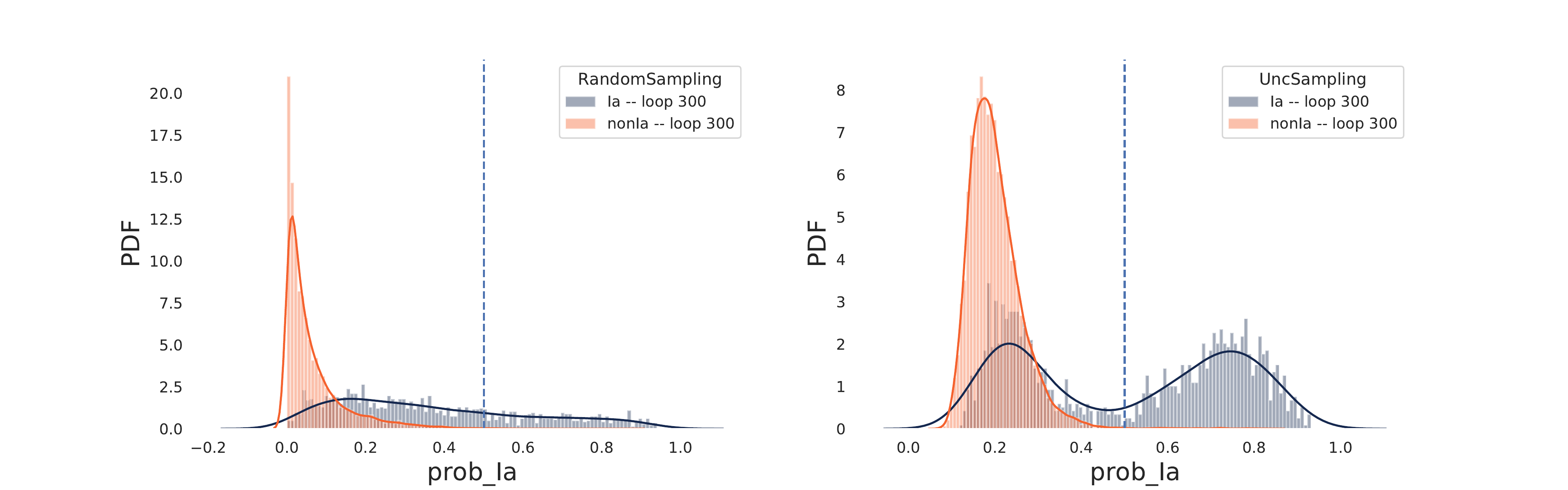}
    \end{minipage}
    \caption{Distribution of classification probabilities in different stages of the learning loop. The top panel shows classification results from the classifier, trained with 10 alerts (5 SNIa and 5 randomly selected from other classes). This initial state was allowed to evolve through 300 iterations following different query strategies. \textbf{Bottom left}: results from a training sample built following a random sampling strategy. \textbf{Bottom right}: results from a training sample built following an uncertainty sampling strategy. The vertical dotted line in each panel denotes 50\% probability.}
    \label{fig:prob_evol}
\end{figure*}

Figure \ref{fig:prob_evol} illustrates how different  query strategies produce different outcomes even when they start from the same initial state. The top panel shows the distribution of classification probabilities for \texttt{SNIa} (blue) and the remaining classes (orange) obtained from a training set with 10 alerts (5 \texttt{SNIa} and 5 randomly selected from other classes). We evolved this system through 300 loops following the two different strategies, thus resulting in 2 final training samples. The bottom left panel shows classification probabilities generated by the RS strategy and bottom right panel shows results from the US strategy. We can see that the probability distribution for non-Ias are very similar in both lower panels.  However, the larger fraction of \texttt{SNIa} in the training sample gathered by the US strategy (Table \ref{tab:queried}) allows the classifier to better identify a subset of clear separable Ias, thus leading to the bimodal behavior seen in the bottom right panel.

\section{Deployment in Fink}
\label{sec:Fink}

The experiments described above resulted in a set of optimized training samples. We used the one which produced the best performing model (accuracy: 0.97, efficiency:  0.53, purity: 0.91, fom: 0.42) to train a random forest classifier, which was subsequently integrated into the broker. Since May/2020, \fink\ has been  processing ZTF alerts using a previous version of this module and reporting \review{candidates with early SN Ia probability higher than 0.5} to TNS since November/2020. The user can access alerts and the light curves from their corresponding objects through the science portal\footnote{\url{https://fink-portal.org}}.

Following the procedure described in Section \ref{sec:features}, in order for an alert to receive a score from this classifier it is required at least 3 observed epochs in each filter. Thus the number of classifiable alerts is strongly influenced by the telescope observation strategy. In order to minimize the effect of this requirement, for the purpose of feature extraction, we also consider observed epochs which fail the broker's quality cuts, but were included in individual alert history, thus allowing for alerts with at least 1 valid point, among the 3 required by the filtering process (Section \ref{sec:filter}), to be classified. In practice, we found that epochs which fail the broker's quality cuts were due to poor signal-to-noise ratio rather than clear bogus (see \cite{fink} for a discussion on the quality cuts). 
In order to tag an alert as an early supernova Ia candidate, we have a set of 6 criteria\footnote{  \href{https://tinyurl.com/FinkIaFilter}{\url{https://tinyurl.com/FinkIaFilter}}}: (a) Ia probability larger than 50\% from the early supernova Ia module described in this work, (b) Ia probability larger than 50\% from either one of the deep learning classifier based on {\sc SuperNNova} \citep{Moller2020} deployed in \fink, (c) no match with galactic objects from the SIMBAD database, (d) non-bogus probability higher than 50\% from the RealBogus algorithm \citep{2019PASP..131c8002M, 2019MNRAS.489.3582D}, (e) Star/Galaxy classification score from SExtractor above 40\% \citep{sextractor}, and (f) the associated object holds no more than 20 photometric measurements.

Alerts fulfilling the above criteria were advertised to the community via publication in TNS. Over this period (01/November/2020 to 31/October/2021), 809 early SN Ia candidates were reported, from which 535 were spectroscopically confirmed. Among the confirmed set, 459 (86\%)  were confirmed as \texttt{SNIa} while the remaining ones were shown to belong to other SN types\footnote{Except one LBV star.}. 

The reported TNS numbers are in agreement with our results (Table \ref{tab:contamination}). However, one should keep in mind that there is still a human screening involved in these numbers, which cannot be modeled. Every observer has the opportunity to inspect the alert candidate and make their own decision about moving forward with the spectroscopic follow-up. Nevertheless, the agreement between the reported numbers and the results presented in this paper are encouraging. 

In order to allow the use of adaptive learning techniques \textit{on-the-fly}, we developed an automated infrastructure which  identifies alerts that can improve the classifier. This sub-stream will be distributed to the community allowing further improvement on future classifications. We plan to extend this framework to enable automatically retraining the classifier as soon as target labels \review{from spectroscopy} are available in public repositories, like TNS.

\section{Conclusions}
\label{sec:conclusions}

The impact of recent technological development and the rise of the era of ``big data'' has been known and widely discussed in a variety of research fields. This new paradigm is certainly also felt in astronomy, however, there are peculiar aspects of astronomical data which require special attention. Among them is the discrepancy between spectroscopic and photometric resources. If we aim to use real data for training as well as testing, we strongly recommend to adopt automatic learning strategies which can work in the small data regime. 

Active learning (AL) is a viable alternative when the number of spectroscopically classified objects is very limited. Working with \review{catalogued transients observed in} the ZTF alert stream as a case study, our final  model achieves  purity of 89\% and efficiency of 54\% while using  merely 310 alerts for training -- the cited results were obtained from a test sample of \referee{23 465} alerts, \referee{$\approx 1463$} of which were SN Ia. 
\review{Such results, derived from a static data base of alerts, were corroborated on real-time ZTF data. In the last year, we reported promising SNe Ia to the TNS service, from which $86\%$ were found to be SNe Ia from spectroscopic follow-up observations.} We stress that this does not correspond to a complete sample (not all candidates were spectroscopically followed) and that we cannot model the influence of human screening while selecting which candidates will be followed. Nevertheless, it is an encouraging result which can be further scrutinized in a purely automated environments, similar to the one proposed by  \citet{street2018}.

Such fully automated pipelines would allow us to remove the current bias present in spectroscopic samples and enable the construction of small, optimized training samples according to the astronomical object of interest. Moreover, this has the potential to result in interesting classified objects benefiting multiple communities. Among the queried alerts, $\approx 45\%$ were \texttt{SNIa}, and others correspond to alerts which can be easily mistaken by SN Ias. The study of such objects may reveal details about their physical processes as well as their relationship with the Ia class. 

This work represents the first step in the development of an adaptive learning environment which has been in the core design of \fink\ since its conception. Nevertheless, we can imagine similar classifiers being specifically design to target other classes of transients. This would allow us to expand our knowledge about the zoology of astronomical transient sources starting from the classes we know -- thus enabling an optimal scientific exploitation of the large volumes of data expected from the next generation of large telescopes.

\begin{acknowledgements}
We thank Johan Bregeon, Jean-Eric Campagne, Benoit Carry and Maria Pruzhinskaya for comments and discussions on the original draft. This work was developed within the Fink community and made use of the Fink community broker resources. Fink is supported by LSST-France and CNRS/IN2P3. E.E.O.I.  receives financial support from CNRS International Emerging Actions under the project \textit{Real-time analysis of astronomical data for the Legacy Survey of Space and Time} during 2021-2022. M.L. acknowledges financial support from the Paris Saclay University, DSI.
\end{acknowledgements}

\bibliographystyle{aa}
\bibliography{ref}

\onecolumn
\appendix
\section{Features values for Figure \ref{fig:fit}}
\label{app:features}

Table \ref{tab:fit} shows the extracted features values for alerts shown in Figure \ref{fig:fit}.

\begin{table*}
\centering
\begin{tabular}{lllll}
 place in Fig. \ref{fig:fit} & top left &            top right &          bottom left &        bottom  right \Bstrut\\
 \hline
  type            &                   Ia &                  AGN &                  EB* &                RRLyr \Tstrut\\
\texttt{objectId} &        ZTF20acqjmzk &          ZTF18aarybyq &        ZTF18aabtvle  &         ZTF18aaptrep \\
 \texttt{candid} &  1424192755315015009 &  1452439215115010000 &  1277247161915010000 &  1385238050715015010 \\
 $a_g$           &             0.527456 &         -3172.429739 &              4.37929 &             0.224062 \\
 $b_g$           &             0.422352 &          3170.602659 &            -0.089541 &            -7.343463 \\
 $c_g$           &          1846.688226 &          4224.205181 &          3766.113687 &          5782.164818 \\
 snratio$_g$     &             6.089286 &            10.370073 &            10.216985 &            13.058625 \\
 nmse$_g$        &             0.000135 &              0.00016 &             0.000044 &             0.000571 \\
 nrise$_g$       &                  6.0 &                  4.0 &                  9.0 &                  8.0 \\
 $a_r$           &             0.356387 &             4.057071 &             0.004883 &             0.021988 \\
 $b_r$           &            -1.713326 &            -0.604576 &          1076.327583 &           271.873134 \\
 $c_r$           &           1913.34661 &         22509.009715 &       1210968.725619 &       1689710.267757 \\
 nsnratio$_r$    &             6.776962 &            23.731405 &             7.220319 &            11.432526 \\
 nmse$_r$        &             0.000032 &             0.000033 &             0.000376 &             0.001975 \\
 nnrise$_r$      &                    5 &                    6 &                    6 &                    7 \\
 \hline
 \end{tabular}
 \caption{Best-fit parameter values found for the alerts shown in Figure \ref{fig:fit}.}
 \label{tab:fit}
 \end{table*}

\clearpage

\end{document}